\documentclass[aps,twocolumn,showpacs]{revtex4}
\usepackage{graphicx}
\usepackage{amssymb}
\usepackage{bm}
\begin{document}


\newcommand{\re}[1]{(\ref{#1})}
\newcommand{\lab}[1]{\label{#1}}
\newcommand{\ci}[1]{\cite{#1}}
\renewcommand{\baselinestretch}{1.25}
\newcommand{\bfr}{\begin{flushright}}
\newcommand{\bfl}{\begin{flushleft}}
\newcommand{\efl}{\end{flushleft}}
\newcommand{\efr}{\end{flushright}}
\newcommand{\bc}{\begin{center}}
\newcommand{\ec}{\end{center}}
\newcommand{\be}{\begin{equation}}
\newcommand{\ee}{\end{equation}}
\newcommand{\bea}{\begin{eqnarray}}
\newcommand{\eea}{\end{eqnarray}}
\newcommand{\ba}{\begin{array}}
\newcommand{\ea}{\end{array}}
\newcommand{\edc}{\end{document}}
\newcommand{\ul}{\underline}
\newcommand{\ri}{\rightarrow\infty}
\newcommand{\li}{\leftarrow\infty}
\newcommand{\ra}{\rightarrow}
\newcommand{\la}{\leftarrow}
\newcommand{\ds}{\displaystyle}
\newcommand{\dsf}{\displaystyle\frac}
\newcommand{\dt}{\Delta{t}}
\newcommand{\il}{\int\limits}
\newcommand{\pal}{\partial}
\newcommand{\xxx}{{\it{X}}}
\newcommand{\bone}{{\bf 1}}
\newcommand{\gComment}[1]{}
\renewcommand{\gComment}[1]{\textcolor{red}{Gerardo: #1}}

\title{Analytic approach to the ground state energy of charged anyon gases in the
high magnetic field}


\author{B. Abdullaev$^1$, U. R\"{o}ssler$^2$,
and C.-H. Park$^3$ }

\address{$^1$
 Theoretical Physics Dept., Institute of Applied Physics,
Uzbekistan National University,
 Tashkent 700174, Uzbekistan}
\address{$^2$
Institute for Theoretical Physics, University of Regensburg, D-93040
Regensburg, Germany}
\address{$^3$
Research Center for Dielectric and Advanced Matter Physics,
Pusan National University, 30
Jangjeon-dong, Geumjeong-gu, Busan 609-735, Republic of Korea.}

\date{Received \today }


\begin{abstract}
We present analytic formulas for the ground state energy of the two-dimensional (2$D$)
anyon gas in the quantum limit of a perpendicular magnetic field (Landau level filling
factor $\nu_L\le 1$). These formulas, for the cases without and with Coulomb interaction,
are obtained by applying the harmonic potential regularization for vanishing confinement
to the harmonically confined Coulomb anyon gas as in our previous paper for the case
without magnetic field. For the case without Coulomb interaction our analytic expression
is exact. It contains a contribution deriving from the anyon gauge field (characterizing
the fractional statistics by the anyon parameter $\nu$) and depends on $\nu$ and $\nu_L$.
For the case with Coulomb interaction we introduce a function, depending on $\nu$, $\nu_L$
and the density parameter $r_s$, which is determined by fitting to the interpolation formula
of Fano and Ortolani in the fractional quantum Hall regime for spin-polarized fermions
in conjunction with results of Yoshioka for the ground state energy of the 2$D$ Coulomb
boson gas in high magnetic fields. With their dependence on $\nu$, our  formulas apply
not only to fermions ($\nu=1$) but quite generally to anyons ($0\le \nu\le 1$).

\end{abstract}



\maketitle

\newpage

\section{Introduction}
\label{sec1}

The two-dimensional electron gas (2$DEG$) in a perpendicular magnetic
field has become a subject of intense theoretical and experimental
investigations due to the observation of the (integer and fractional)
quantum Hall effects (IQHE, FQHE) \ci{prange} and of the Spin Hall effect
(SHE) \ci{kato,wunderlich} observed in semiconductor quantum structures.
Disorder and Coulomb interaction are responsible for the quantum Hall effects, of which
the IQHE is dominated by disorder, while the FQHE is dominated by correlation due to
Coulomb interaction \ci{laupr}, which for fractional fillings of the
lowest Landau level leads to a liquid state of quasi-particles obeying fractional
statistics (or having fractional charges) \ci{fis}.
For the SHE, the role of a weak current-induced magnetic field for the
polarization of spins at the edges of quasi 2$D$ semiconductors is still
unclear \ci{choi}. One can expect that these
effects are only first observations of interesting properties of 2$DEG$es,
in particular of the spin-polarized systems, exposed to a magnetic field, which
may uncover new aspects of the structure of condensed matter. Another object --
the 2$D$ Coulomb Bose gas may attract attention for the explanation of
High-$T_c$ superconductivity \ci{abdullaev1}. Among the properties of these
systems the ground state energy and its dependence on the system parameters is
of principal interest.

Calculations of the ground state energy of 2$D$ electrons in a magnetic field
reported in the literature are devoted mainly to the FQHE regime. In his
pioneering work, Laughlin \ci{laughlin} demonstrated that in the strong magnetic
field the energy of spin-polarized electrons can be lower
than that of a charge-density wave state in the lowest Landau level. By assuming
no mixing of Landau levels, he proposed an approximate expression for the ground
state energy. By using Monte Carlo simulations for a classical one-component $2D$
plasma, Levesque et al. \ci{levesque} obtained a more accurate result than Laughlin.
Fano and Ortolani \ci{fano} corrected this result close to the Landau level
filling factor $\nu_L\approx1$ by taking into account the
electron-hole symmetry. Quantum Monte Carlo calculations of the ground state
energy were performed for fixed values of $\nu_L<1$ and of the density parameter
$r_s$ in \ci{ortiz}
and \ci{price}. With these numerical data an interpolation
formula was found for the interval $0\le r_s\le100 \,$ \ci{price}. The
ground state energy for the Wigner crystal in a strong magnetic field
was obtained by Variational Monte Carlo calculations \ci{zhu}. All these
papers provide information on the ground state energy of
the spin-polarized $2DEG$ in strong magnetic fields and for discrete values of 
$r_s$. It would be desirable, however, to have an analytic (though
approximate) formula for the ground state energy that covers the whole range of
the system parameters and applies also to particles obeying different statistics.

In a previous paper \ci{arm}, we have derived an approximate
analytic formula for the ground state energy of the 2$D$ Coulomb
anyon gas without magnetic field. It was obtained from the corresponding result
for the harmonically confined 2$D$ Coulomb anyon gas by applying
a regularization procedure for vanishing confinement \ci{aormn}. In order 
to account for the fractional statistics and Coulomb interaction, we
introduced a function, which depends on both the statistics and
density parameters ($\nu$ and $r_s$, respectively), and determined
this function by fitting to the ground state energies of the
classical $2D$ electron crystal at very large $r_s$ (the 2$D$ Wigner
crystal) and of the dense 2$D$ Coulomb Bose and Fermi gases at very small $r_s$.
Applied to the spin-polarized electron ($\nu=1$) and charged boson ($\nu=0$) cases,
our analytic expression yields ground state energies, which are in reasonable
agreement with numerical as well as analytic data provided in the
literature.

For the harmonically confined anyons \ci{aormn}, we have already considered
the effect of a magnetic field. In the present paper, we apply the vanishing
confinement regularization procedure to these results, thus extending the
calculation of \ci{arm} to the magnetic field case. The formula derived in
\ci{aormn} for the ground state energy of the confined system (including the
magnetic field) differs from the field-free case by a parameter, which contains the
ratio $\omega_0^2/\omega_c^2$ as outlined already in \ci{wensauer}, where $\omega_0$
and $\omega_c$ are the $2D$ oscillator and cyclotron frequencies, respectively.
Taking into account that the strong magnetic field condition for $\omega_c$ is
determined in Ref.~\ci{aormn} for fixed numbers of fermions ($\nu=1$) and going to the
thermodynamic limit, $\omega_0/\omega_c$  becomes the Landau level filling factor
of fermions $\nu_L=2 l_H^2/r_0^2$, where $l_H$ is the magnetic length and
$r_0=r_s a_B$ ($a_B$ being the Bohr radius). In the limit of strong
magnetic fields ($\nu_L\leq 1$), we obtain the analytic expression for the ground
state energy per particle for the cases without and with Coulomb interaction.
While for the case without interaction it is exact and depends on the statistical
parameter $\nu$, $\nu_L$ and $r_s$, it is approximate for the case with interaction 
and depends on some function of these parameters, as will be shown here. We determine
this function by fitting to the interpolation formula of Fano and Ortolani for the
fractional quantum Hall states \ci{fano}, which will be modified by taking into
account the numerical result of Yoshioka \ci{yoshioka} for the ground state energy
of 2$D$ Coulomb boson gas in the high magnetic field. The dependence on the parameter
$\nu$ means to generalize the result of Ref.~\ci{fano} for spin-polarized electrons to anyons.

In the Sec. \ref{sec2} we describe and apply the idea of the harmonic potential regularization
procedure to get, in the thermodynamic limit, the analytic expression of the ground state energy
per particle yet for noninteracting anyons in a magnetic field. The expression of this energy
for the system with Coulomb interaction will be derived in the Sec. \ref{sec3} and we summarize
and conclude the paper in the last section.

\section{Noninteracting case} \label{sec2}

Let us start with Eq.~(28) of our previous work \ci{aormn}
\be
E_0=N\hbar \left (\omega_0^2 +\dsf{\omega _c^2 } {4}\right )^{1/2}{\cal
N}^{1/2} - \dsf{\nu \beta \hbar \omega _c}{4}N(N-1) \,
\lab{agmf1}
\ee
for the minimum energy of $N$ anyons in a parabolic confinement and a magnetic field. In
\ci{aormn} we have specified the expression for ${\cal N}$ for the cases of weak and strong
magnetic field, for which the analytical dependence of $E_0$ on $N$ can be derived. For weak
magnetic fields, $0\leq \omega _c \leq \omega_0/(K(K-1))^{1/2}$ and $K\geq2$ being the number
of closed shells of anyon quantum states in a harmonic potential, we had ${\cal N}=1+|\nu| (N-1)$,
while for strong magnetic fields, $\omega _c \geq \omega_0 (N-2)/(N-1)^{1/2}$, (obtained
for fermions) we had ${\cal N} =(1+ \nu\beta(N-1)/2)^2$. From \ci{arm} we know already that the
harmonic potential regularization requires for the case without magnetic field  $\omega_0\sim 1/N^{1/2}$.
In addition, the number $K$ of closed shells increases with the number of particles $N$ to
infinity. Thus the interval for the weak magnetic field case shrinks to zero and no (analytic)
dependence of the ground state energy per particle, ${\cal E}_0=E_0/N$, on the magnetic field can be
obtained.

Making use of the invariance of the energy expression, Eq.~\re{agmf1}, under simultaneous sign changes of
$\beta =eH/|eH|$ (the magnetic field direction) and of the anyon parameter $\nu =e\phi/(2\pi\phi_0)$
(see also Ref.~\ci{ouvry}) (where $\phi $ is the magnetic flux of the anyon gauge field and $\phi_0=e/hc$ is the elementary
flux quantum) we may replace $\beta \nu$ by $\nu$ with $0\le \nu\le 1$. Thus the expression for
${\cal N}$ for the case of strong magnetic fields reads ${\cal N} =(1+ \nu(N-1)/2)^2$ and in
the second term of Eq.~\re{agmf1} we write $\nu$ instead of $\nu \beta$. Starting from this formula,
we apply the harmonic potential regularization procedure of \ci{arm} to obtain an approximate analytic
expression for ${\cal E}_0$ in the thermodynamic limit ($N\rightarrow \infty$) for noninteracting
anyons.

For the strong magnetic field case, the harmonic potential regularization leads with
$\omega_0\sim 1/N^{1/2}$ to a lower limit of the cyclotron frequency (or magnetic field), which
for $N\rightarrow \infty$ becomes independent of the particle number $N$. For this case, the ground
state energy per particle takes the form
\be
{\cal E}_0=\hbar \left (\omega_0^2 +\dsf{\omega _c^2 }{4}\right )^{1/2}\left(1+ \dsf{\nu(N-1)}{2}\right ) -
\dsf{\nu \hbar \omega _c}{4}(N-1) \, .
\lab{agmf2}
\ee

In analogy with Ref.~\ci{arm}, we introduce for the noninteracting case $\omega_0=\hbar f/(Mr_0^2N^{1/2})$,
where now $f$ is an unknown constant. Thus in the limit $N\rightarrow \infty$ the strong magnetic field case
is defined by  $\hbar\omega _c\geq \hbar^2 f/(Mr_0^2)$. Keeping in mind that $\hbar\omega _c= \hbar^2 /(Ml_H^2)$,
this relation reduces to $f l_H^2/r_0^2\leq 1$, where $f l_H^2/r_0^2$ can be understood
as the Landau level filling factor for  spin-polarized fermions  ($\nu=1$) with $f=2$ (we recall here that
the strong magnetic field condition was obtained for fermions), i.e., $\nu_L=2l_H^2/r_0^2$.

Expanding the first term of Eq.~\re{agmf2} in powers of $4\omega_0^2/\omega _c^2$ and substituting
$\omega_0$ we obtain in the thermodynamic limit ($N\rightarrow \infty$) the exact expression for
the ground state energy per particle, which is now written in terms of $\nu$, $\nu_L$ and $r_s$
and in proper energy units ($Ry=Me^4/(2\hbar^2)$),
\be
{\cal E}_0(\nu,\nu_L,r_s)-\dsf{2}{\nu_L r_s^2}= \dsf{2\nu \nu_L}{r_s^2} \, .
\lab{agmf3}
\ee
We note that ${\cal E}_0(\nu,\nu_L,r_s)$ is independent of the particle number $N$. The term, which is subtracted
on the left hand side, is the energy of the lowest Landau level, $\hbar \omega _c/2$, expressed in $Ry$ energy units.
The term on the right hand side, depending on $\nu$ and $\nu_L$, vanishes in both cases when we either decrease the anyon
parameter $\nu$ or increase the magnetic field.

Let us recall here the corresponding result for the case without magnetic field from our previous work
\ci{arm}, where the harmonic potential regularization has led to the ground state energy
\be
E_0(N\rightarrow \infty, \nu)=\pi \hbar^2 \nu \rho N/M
\lab{agmf4p}
\ee
of $N$ anyons (without Coulomb interaction) in the thermodynamic limit ($N\rightarrow \infty$). For spin-polarized fermions ($\nu=1$)
this is the energy of a filled Seitz circle with radius $2^{1/2} k_F,\,$ $k_F=(2\pi \rho)^{1/2}$ and the
areal particle density $\rho$. Thus Eq.~\re{agmf4p} describes the ground state energy of anyons, which
interact only via the anyon gauge field. In the light of this result, we understand ${\cal E}_0(\nu,\nu_L,r_s)$
(Eq.~\re{agmf3}) as the ground state energy per anyon in the presence of an external magnetic field and
the right hand side of Eq.~\re{agmf3} as contribution due to the anyon gauge field (or else the Pauli principle).
To our knowledge, this result is not yet reported in the literature.

\section{Coulomb interacting case} \label{sec3}

In order to include the Coulomb interaction, we start from Eq.~(50) of Ref.~\ci{aormn}, which we write
in the form
\be
{\cal E}_0=\dsf{E_0}{N}=\hbar \omega_0 c^{1/3} \left[\dsf{{\cal N} c^{1/3}}{2\bar{X}_0^2}+\dsf{\bar{X}_0^2}{2}+
\dsf{{\cal M}}{\bar{X}_0} \right] - \dsf{\nu \hbar  \omega _c (N-1)}{4} \,
\lab{agmf5}
\ee
with
\be
\bar{X}_0=(\bar{A}+\bar{B})^{1/2}+[-(\bar{A}+\bar{B})+2(\bar{A}^2-\bar{A}\bar{B}+\bar{B}^2)^{1/2}]^{1/2} \ ,
\lab{agmf6}
\ee
where
\bea
\ba{l}
\bar{A}=\left[{\cal M}^2/128+\left((c {\cal N}/12)^3+({\cal M}^2/128)^2
\right)^{1/2} \right]^{1/3},\\
\bar{B}=\left[{\cal M}^2/128-\left((c {\cal N}/12)^3+({\cal M}^2/128)^2
\right)^{1/2} \right]^{1/3}.
\lab{agmf7}
\ea
\eea
Note that we have replaced $\nu \beta$ by $\nu$ (see Sec. \ref{sec2}) and $1+\omega_c^2/(4\omega_0^2)$ by $c$.

In Eqs. \re{agmf5} and \re{agmf7} we use from Ref.~\ci{arm} (Eq.~(32) with $L=(\hbar /M\omega_0)^{1/2}$)
\be
{\cal M}=-\dsf{a N^{1/2}}{a_B}\left( \dsf{\hbar}{M\omega_0}\right)^{1/2} \, .
\lab{agmf8}
\ee
While in Ref.~\ci{arm} $a$ has been introduced as a constant, it will become now a
function of $\nu$, $\nu_L$ and $r_s$.
In the strong magnetic field case, $\nu_L\leq 1$, we use
again ${\cal N} =(1+ \nu(N-1)/2)^2$ and perform the harmonic potential regularization to
reach the thermodynamic limit $N\rightarrow \infty$ (using as in the previous section
$\omega_0=\hbar f/(Ma_B^2r_s^2 N^{1/2})$ with $f=2$) and the 2$D$ system.

In the limit $N\rightarrow \infty$, we may expand the right
hand side of Eq.~\re{agmf5} (with $\bar{X}_0$ from Eq.~\re{agmf6} and $\bar{A}$ and $\bar{B}$ from Eq.~\re{agmf7})
in the small parameter $t=({\cal M}^2/128)^2/(c {\cal N}/12)^3)$ by using ${\cal M}$, $\omega_0$ and ${\cal N}$
as described before. Up to first order in $t$ this expansion reads
\be
{\cal E}_0=\hbar \omega_0 c^{1/3} \left( {\cal N}^{1/2}c^{1/6}-\dsf{|{\cal M}|}{{\cal N}^{1/4}c^{1/12}}\right)-
\dsf{\nu \hbar  \omega _c (N-1)}{4} \, ,
\lab{agmf9}
\ee
and provides for $N\rightarrow \infty$ the expression (in $Ry$ energy units)
\be
{\cal E}_0(\nu,\nu_L,r_s)-\dsf{2}{\nu_L r_s^2}= \dsf{2\nu \nu_L}{r_s^2}- \dsf{2^{3/2}a(\nu,\nu_L,r_s)}{r_s
(\nu \nu_L)^{1/2}}\,
\lab{agmf11}
\ee
for the ground state energy per particle in the case with Coulomb interaction.
All higher order terms in $t$ on the right hand side carry powers of $-|{\cal M}|/(c^{1/4}{\cal N}^{3/4})$
as factors and vanish for $N\rightarrow \infty$. Therefore, Eq.~\re{agmf11} is exact in the thermodynamic limit.
According to Eq.~\re{agmf11}, the ground state energy per anyon in the strong magnetic field case contains besides the
energy of free particles $\hbar \omega _c/2=2/(\nu_L r_s^2)$ in the lowest Landau level two terms on the
right hand side, of which the first one is known from Eq.~\re{agmf3}, while the second one is due to Coulomb and
statistical interaction. So far, as in the noninteracting case (Eq.~\re{agmf3}) the ground state energy of the interacting
anyons depends on $\nu$, $\nu_L$ and $r_s$.

To determine $a(\nu,\nu_L,r_s)$ we use the Fano and Ortolani formula \ci{fano} for the correlation energies of
different FQHE states of spin-polarized fermions (in $Ry$ energy units)
\begin{eqnarray}
E_c^{FO}(\nu_L,r_s)&=& \dsf{2\sqrt{2}}{r_s}[-0.782133(1-\nu_L)^{3/2} \nonumber\\
&+&(0.683(1-\nu_L)^2-(\pi/8)^{1/2})\nu_L^{1/2} \nonumber\\
&-&0.806\nu_L(1-\nu_L)^{5/2}] \,
\lab{agmf12}
\end{eqnarray}

and note that the left hand side of Eq.~\re{agmf11} should obey the following constraints:

i) In the fermion case ($\nu=1$), it  should be equal to the interpolating formula, Eq.~\re{agmf12}.

ii) For vanishing Landau level filling factor ($\nu_L\rightarrow 0$) it should become independent of the anyon
parameter $\nu$, because this limit corresponds to the classical 2$D$ Wigner crystal.

iii) For the boson gas ($\nu=0$), it should reproduce the numerical results of Yoshioka (see Fig. 3(b) of \ci{yoshioka})
for the 2$D$ Coulomb boson gas in the lowest Landau level.

As it turns out, the results of Yoshioka (except for the cusps at fractional fillings) can be interpolated by
modifying the Fano and Ortolani formula, Eq.~\re{agmf12}. This is achieved by replacing (inside of the square brackets)
the second term by $(0.683\nu (1-\nu_L)^2-(A(1-\nu)+\nu(\pi/8)^{1/2}))\nu_L^{1/2}$, where $A=0.45$, and the third term by
$0.806(\nu+(1-\nu)\nu_L)\nu_L(1-\nu_L)^{5/2}$. If we denote the modified expression of $E_c^{FO}(\nu_L,r_s)$ as
$E_c(\nu,\nu_L,r_s)$ then, by equating the right hand side of the Eq.~\re{agmf11} with $E_c(\nu,\nu_L,r_s)$, one obtains
\be
a(\nu,\bar{\nu},r_s)=\dsf{r_s (\nu \nu_L)^{1/2}}{2^{3/2}}\left(\dsf{2\nu \nu_L}{r_s^2}-E_c(\nu,\nu_L,r_s)\right)
\, .
\lab{agmf13}
\ee

We express the ground state energy per particle in $e^2/l_H$ energy units. This means in Eq.~\re{agmf12} to
replace the common factor $2\sqrt{2}/r_s$ by $\nu_L^{1/2}$ and to change the notation of energies from  $E_c^{FO}(\nu_L,r_s)$
and $E_c(\nu,\nu_L,r_s)$ into $E_c^{FO}(\nu_L)$ and $E_c(\nu,\nu_L)$, respectively. In Fig. 1 we show the dependence of
the correlation energy ${\cal E}_0(\nu,\nu_L)-\hbar\omega _c/2$ (expressed in $e^2/l_H$ energy units)
as function of $\nu_L$ for two limiting cases of charged anyon gases: spin-polarized fermions  ($\nu=1$), for which
$E_c(\nu=1,\nu_L)=E_c^{FO}(\nu_L)$, and bosons ($\nu=0$), for which the correlation energy is $E_c(\nu=0,\nu_L)$.
The two curves shown represent approximate interpolations of the results for fermions and bosons in Ref. \ci{yoshioka}
but extend these results to the whole range of filling factors $0\le \nu_L\le 1$.

\begin{figure}
\begin{center}
\includegraphics[angle=270,width=9.5cm,scale=1.0]{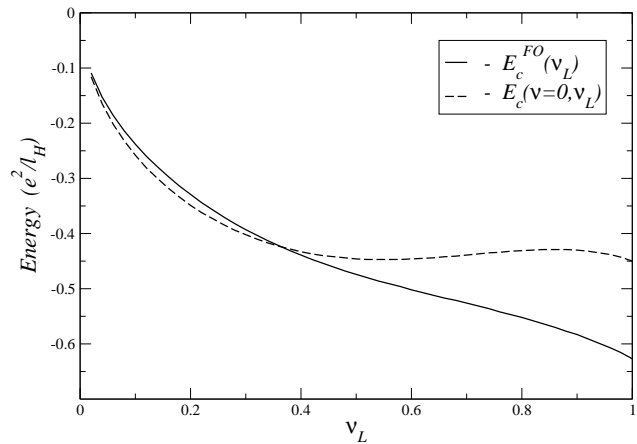}
\end{center}
\caption{Correlation energy per particle in the lowest Landau level (expressed in $e^2/l_H$ energy units)
vs. Landau level filling factor $\nu_L$ for spin-polarized fermions, $E_c(\nu=1,\nu_L)=E_c^{FO}(\nu_L)$, (solid line)
and for bosons, $E_c(\nu=0,\nu_L)$, (dashed line).
} \lab{fig1}
\end{figure}

Thus Eq.~\re{agmf11} together with Eq.~\re{agmf13} represents the analytic formula for the ground state energy
of charged anyons in the quantum limit of an applied magnetic field. By comparing with the corresponding result for
free anyons (see Eq.~\re{agmf3} in Sec. \ref{sec2}), we can ascribe the second term on the right hand side
of Eq.~\re{agmf11} to the Coulomb interaction in the presence of a high magnetic field.

\section{Conclusions}
\label{sec4}

We have derived analytic formulas for the ground state energy of the Coulomb anyon gas in a perpendicular
strong magnetic field. Following the concept of our previous paper for the case without magnetic field, we
applied the harmonic potential regularization to the harmonically confined 2$D$ Coulomb anyon gas in the
magnetic field. This concept is based on flattening out the confinement potential with simultaneously
increasing of the particle number to reach the thermodynamic limit. For the noninteracting anyon system
we derive an analytic expression for the ground state energy per particle, which applies to the high magnetic
fields and is exact in the thermodynamic limit. It contains a contribution from the anyon gauge potential,
which was not known so far. For the interacting anyon system our analytic energy expression contains an
unknown function, which is determined by fitting to the modified interpolation formula of Fano and Ortolani,
which takes into account the ground state energy of 2$D$ Coulomb boson gas in the high magnetic field.

\section{Acknowledgements}

The activity of B. A. has been supported by the Volkswagen Foundation and he
thanks for the hospitality at the University of Regensburg.


\begin{thebibliography}{99}

\bibitem{prange}
{\it The Quantum Hall Effect}, edited by R. E. Prange and S. M.
Girvin (Springer-Verlag, Berlin, 1987).

\bibitem{kato}
Y. K. Kato, R. C. Myers, A. C. Gossard, and D. D. Awschalom, Science
{\bf 306}, 1910 (2004).

\bibitem{wunderlich}
J. Wunderlich, B. Kaestner, J. Sinova, and T. Jungwirth, Phys. Rev.
Lett. {\bf 94}, 047204 (2005).

\bibitem{laupr}
See article of R. B. Laughlin in \ci{prange}.

\bibitem{fis}
S. Forte,  Rev. Mod. Phys. {\bf 64}, 193 (1992); R. Iengo and K.
Lechner,  Phys. Rep. {\bf 213}, 179 (1992); M. Stone (editor), {\it
Quantum Hall Effect} (World Scientific, Singapore, 1992).

\bibitem{choi}
B. Abdullaev, M. Choi, and C. -H. Park, J. Korean Phys. Soc., {\bf 49}, S638 (2006).

\bibitem{abdullaev1}
B. Abdullaev, {\it Trends in Boson Research}, edited by
A. V. Ling (Nova Science Publishers, N. Y., 2006), pp 139-161; B. Abdullaev
and C. -H. Park, J. Korean Phys. Soc., {\bf 49}, S642 (2006).

\bibitem{laughlin}
R. B. Laughlin, Phys. Rev. Lett. {\bf 50}, 1395 (1983).

\bibitem{levesque}
D. Levesque, J. J. Weis, and A. H. MacDonald,  Phys. Rev. B {\bf
30}, 1056 (1984).

\bibitem{fano}
G. Fano and F. Ortolani,  Phys. Rev. B {\bf 37}, 8179 (1988).

\bibitem{ortiz}
G. Ortiz, D. M. Ceperley, and R. M. Martin, Phys. Rev. Lett. {\bf
71}, 2777 (1993).

\bibitem{price}
R. Price, P. M. Platzman, and S. He, Phys. Rev. Lett. {\bf 70}, 339
(1993).

\bibitem{zhu}
X. Zhu and S. G. Louie, Phys. Rev. Lett. {\bf 70}, 335 (1993).

\bibitem{arm}
B. Abdullaev, U. R\"{o}ssler, and M. Musakhanov, Phys. Rev. B {\bf 76}, 075403 (2007).

\bibitem{aormn}
B. Abdullaev, G. Ortiz, U. R\"{o}ssler, M. Musakhanov, and A. Nakamura,
Phys. Rev. B {\bf 68}, 165105 (2003).

\bibitem{wensauer}
A. Wensauer and U. R\"{o}ssler, Phys. Rev. B{\bf 69}, 155302 (2004).

\bibitem{yoshioka}
D. Yoshioka, Phys. Rev. B{\bf 29}, 6833 (1984).

\bibitem{ouvry}
A. Dasnieres de Veigy and S. Ouvry, Phys. Rev. Lett. {\bf 72}, 600 (1994).

\bibitem{bm}
L. Bonsal and A. A. Maradudin, Phys. Rev. B {\bf 15}, 1959 (1977).

\end{thebibliography}
\end{document}